\documentclass[aps,prb,preprint,showpacs]{revtex4}
\usepackage{natbib}
\usepackage{epsfig}
\usepackage{color}

\begin{document}

\title{Itinerant surfaces with spin-orbit couplings, correlations and
external magnetic fields: Exact results}
\author{N\'ora~Kucska and Zsolt~Gul\'acsi}%$^{a,b}$ and Dieter~Vollhardt$^{a}$}
\affiliation{Department of Theoretical Physics, University of
Debrecen, H-4010 Debrecen, Bem ter 18/B, Hungary}
%$^{(a)}$ Theoretical Physics III, Center for
%Electronic Correlations and Magnetism, Institute for Physics,
%University of Augsburg, D-86135 Augsburg,
%Germany \\
%$^{(b)}$ Department of Theoretical Physics, University of
%Debrecen, H-4010 Debrecen, Hungary}
\date{\today }
%\date{September 24, 2015}

\begin{abstract}
We analyze, in exact terms,
multiband 2D itinerant correlated fermionic systems 
with many-body spin-orbit interactions, and in-plane external 
magnetic fields. Even if such systems with broad applicability in leading 
technologies are non-integrable, we set up an exact solution 
procedure for them, which is described in details. Casting the 
Hamiltonian in positive semidefinite form, the technique leads to the 
ground state, and also characterizes the low lying excitation spectrum.
\end{abstract}

%\pacs{PACS No. 71.10.Fd, 71.10.Hf, 71.10 Pm, 71.70.Ej, 05.30.Fk, 67.40.Db}

\maketitle

\section{Introduction}

Surfaces with spin-orbit interactions (SOI) are the subject for a broad area of 
current research (see the review \cite{0N19} ), SOI providing essential effects
in various phenomena of large interest today, ranging from quantum magnets
\cite{1N19}, topological phases \cite{2N19}, ultracold atom experiments 
\cite{3N19}, to Majorana fermions \cite{4N19}. The applications appear mostly 
in low dimensional systems \cite{5N19,6N19,7N19,8N19,9N19,10N19,11N19}, and 
during processing, often external fields are as well present, the most 
interesting applications being related to strongly correlated systems.
Contrary to its importance, although exact treatments of 2D strongly correlated systems 
with spin-orbit coupling are being developed \cite{11N19},
studies including applied external magnetic fields are absent. Our aim in this 
Letter is to fill up this gap by setting up the details of a calculation 
procedure for such situations, considering Hamiltonians describing 
realistic correlated systems.

The main difficulty encountered is that the here studied
2D systems are non-integrable, so special techniques must be used in order to
describe them in exact terms. For this reason we use the method based on 
positive semidefinite operator properties whose applicability does not 
depend on dimensionality and integrability \cite{T1,T2,T3,T4}. The method has been previously applied
in conditions unimaginable before
in the context of exact solutions in 1-3D, even in the presence of the
disorder \cite{T5,T6,T7,T8,T9,T10,T11,T12}.

\section{The system analysed}

The Hamiltonian of the system has the form 
$\hat H=\hat H_{kin}+\hat H_{int}+ \hat H_h$, 
\begin{eqnarray}
\hat H = \sum_{p,p'} \sum_{{\bf i},{\bf r}} \sum_{\sigma,\sigma'} (k^{p,p';\sigma,\sigma'}_{
{\bf i},{\bf i}+{\bf r}} \hat c^{\dagger}_{p,{\bf i},\sigma} \hat c_{p',{\bf i}+
{\bf r},\sigma'} + H.c.) +
\sum_{p} \sum_{{\bf i}} U_{p,{\bf i}} \hat n_{p,{\bf i},\uparrow} 
\hat n_{p,{\bf i},\downarrow} + \sum_{p,{\bf i}} \sum_{\sigma,\sigma'} {\vec h_{p,{\bf i}}}
\hat c^{\dagger}_{p,{\bf i},\sigma}{\vec \sigma_{\sigma,\sigma'}} \hat c_{p,{\bf i},\sigma'}.
\label{Equ1}
\end{eqnarray}
where the first term represents the kinetic part of the Hamiltonian
($\hat H_{kin}$), the second term is the interaction part
($\hat H_{int}$), while the last term describes the interaction with the
external magnetic field ($\hat H_h$). At the level of $\hat H_{kin}$,
in order to have a realistic 2D surface description, two bands are considered,
denoted hereafter by p,p'=a,b. However we note,
that this choice not diminishes the applicability of the deduced results, since
usually, the theoretical description of muliband systems is given by projecting
the multiband structure in a few-band picture \cite{RDA}, projection which is 
stopped here only for its relative simplicity at two-bands level. Again in 
order to approach a real systems, besides on-site one particle terms
(${\bf r}=0$), one takes into consideration nearest-neighbor (${\bf r}=
{\bf x}_1,{\bf x}_2$, where ${\bf x}_1,{\bf x}_2$ are the Bravais vectors), 
and next nearest-neighbor (${\bf r}= {\bf x}_2+{\bf x}_1,{\bf x}_2-
{\bf x}_1$) contributions. Furthermore, note that the $k^{p,p';\sigma,\sigma'}_{
{\bf i},{\bf i}+{\bf r}}$ coefficient represents for $(p=p',{\bf r}=0)$, 
$(p=p',{\bf r} \ne 0)$ on-site potential, (hopping matrix element); while for
$(p \ne p',{\bf r}=0)$, $(p \ne p',{\bf r} \ne 0)$ on-site hybridization,
(inter-site hybridization).
Concerning $\hat H_{int}$, since in itinerant 
many-body systems strong screening effects are present, we consider at this 
stage only the on-site Coulomb repulsion (Hubbard interaction term) in the 
correlated band (p=b, $U_b >0$), the second band being considered 
non-correlated (p=a, $U_a=0$). 
The many-body spin-orbit interactions being of one-particle 
type, are introduced in the kinetic part of the Hamiltonian, explicitly 
in the nearest neighbor spin-flip hopping terms, i.e. 
coefficients $k^{p,p;\sigma, -\sigma}_{{\bf i},{\bf i}+{\bf r}}, \: {\bf r}={\bf x}_1,
{\bf x}_2$. These terms are of Rashba
($\lambda^p_R, \: p=a,b$) and Dresselhaus ($\lambda^p_D, \: p=a,b$) type 
\cite{RD}. Consequently, one has for ${\bf r}={\bf x}_1$, the structure
$k^{p,p;\uparrow,\downarrow}_{{\bf i},{\bf i}+{\bf x}_1} = \lambda^p_R - i \lambda^p_D$,
$k^{p,p;\downarrow,\uparrow}_{{\bf i},{\bf i}+{\bf x}_1} = -\lambda^p_R - i \lambda^p_D$,
while for  ${\bf r}={\bf x}_2$ the expressions
$k^{p,p;\uparrow,\downarrow}_{{\bf i},{\bf i}+{\bf x}_2} = \lambda^p_D - i \lambda^p_R$,
$k^{p,p;\downarrow,\uparrow}_{{\bf i},{\bf i}+{\bf x}_2} = -\lambda^p_D - i \lambda^p_R$.
We underline that 
even if usually the SOI contributions are small, 
they introduce essential effects since they
break the double spin-projection degeneracy of each band.
Hence, in the presence of strong correlations,
the essential effects introduced cannot be obtained by standard perturbation 
approximations \cite{11N19}. We note that other spin-flip terms
are not present in $\hat H_{kin}$, and one has for all considered ${\bf r}$
values $k^{p,p';\uparrow,\uparrow}_{{\bf i},{\bf i}+{\bf r}}=k^{p,p';\downarrow,\downarrow}_{
{\bf i},{\bf i}+{\bf r}}=k^{p,p'}_{{\bf i},{\bf i}+{\bf r}}$. Furthermore, 
in order to not diminish the effect of the spin-flip nearest-neighbor
hopping terms produced by SOI, the external fields are only applied in-plane,
hence without the z-component ($h^z_{p,{\bf i}}=0, h^x_{p,{\bf i}}, h^y_{p,{\bf i}}
\ne 0$). We underline, that the in-plane
$h^x_{p,{\bf i}}, h^y_{p,{\bf i}}$ contributions will additively renormalize the
$k^{p,p;\sigma,-\sigma}_{{\bf i},{\bf i}}$ contributions as 
$\bar k^{p,p;\uparrow,\downarrow}_{{\bf i},{\bf i}}= k^{p,p;\uparrow,\downarrow}_{{\bf i},{\bf i}} +
h^x-ih^y, \: (\bar k^{p,p;\downarrow,\uparrow}_{{\bf i},{\bf i}})^* = 
\bar k^{p,p;\uparrow,\downarrow}_{{\bf i},{\bf i}}$.
%%%%%%%%%%%%%%%%%%%%%%%%%%%%%%%%%%%%%%%%%%%%%%%%%%%%%%%%%%%%%%%%%%%%%%%%%%%%%
% FIGURE 1
%%%%%%%%%%%%%%%%%%%%%%%%%%%%%%%%%%%%%%%%%%%%%%%%%%%%%%%%%%%%%%%%%%%%%%%%%%%%%
\begin{figure}[h]
\centerline{\includegraphics[width=4cm,height=4cm]{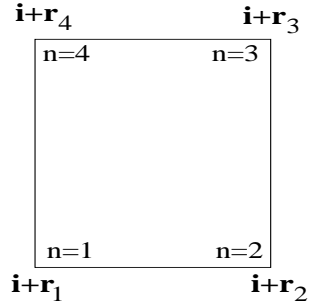}} 
%\centerline{\epsfbox{NoraBandsFig1u.eps}}
\caption{Unit cell defined at the lattice site ${\bf i}$ with in-cell
notations of sites $n=1,2,3,4$.} 
\label{fig1}
\end{figure}
%%%%%%%%%%%%%%%%%%%%%%%%%%%%%%%%%%%%%%%%%%%%%%%%%%%%%%%%%%%%%%%%%%%%%%%%%%%%%

\section{The Hamiltonian cast in pozitive 
semidefinite form}

\subsection{The transformation of the Hamiltonian}

Now we
turn back to (\ref{Equ1}), and present the transformation of $\hat H$
in exact terms. On this line
we introduce two block operators Q=A,B for each site ${\bf i}$, which for a
fixed Q value are defined as
\begin{eqnarray}
\hat Q_{\bf i} = \sum_{p=a,b} \sum_{n=1,2,3,4} \sum_{\alpha=\uparrow,\downarrow}
q_{Q,p,n,\alpha} \hat c_{p,{\bf i}+{\bf r}_n,\alpha}.
\label{Equ2}
\end{eqnarray}
Here, in order ${\bf r}_1=0, \: {\bf r}_2={\bf x}_1, \: {\bf r}_3={\bf x}_1+
{\bf x}_2$, and ${\bf r_4}={\bf x}_2$, see Fig.1. At a given lattice 
site ${\bf i}$, for a fixed Q and p
value, the $\hat Q_{\bf i}$ operator has 8 contributions, 4 for spin $\alpha=
\uparrow$, and other 4 for spin $\alpha=\downarrow$. For fixed $\alpha$ the
mentioned 4 values denoted by $n=1,2,3,4$ are placed in the four corners of an
elementary plaquette connected to the lattice site ${\bf i}$.
Using (\ref{Equ2}), the starting system Hamiltonian $\hat H$ in (\ref{Equ1}) 
becomes of the form  
\begin{eqnarray}
\hat H = \hat P + S_c, 
\label{Equ2a}
\end{eqnarray}
where $\hat P$ represents a 
positive semidefinite operator, while $S_c$ a scalar. Taking into account
that $\hat P = \hat P_Q + \hat P_U$ where $\hat P_U = U_b \sum_{\bf i} \hat P_{
\bf i}$, for $U_b > 0$ one has
\begin{eqnarray}
&&\hat P_Q = \sum_{\bf i}\sum_{Q=A,B} \hat Q_{\bf i} \hat Q^{\dagger}_{\bf i}, \quad
\hat P_{\bf i} =
\hat n_{b,{\bf i},\uparrow} \hat n_{b,{\bf i},\downarrow} -(\hat n_{b,{\bf i},\uparrow} + 
\hat n_{b,{\bf i},\downarrow}) +1,
\nonumber\\
&&S_c = \eta N -U_b N_{\Lambda} - \sum_{\bf i}\sum_{Q=A,B} d_{{\bf i},Q},
\quad d_{{\bf i},Q}=\{\hat Q_{\bf i}, \hat Q^{\dagger}_{\bf i} \}, 
\label{Equ3}
\end{eqnarray}
where $N (N_{\Lambda})$ represents the number of electrons (lattice sites).

The corresponding matching equations which allows the transformation of the
starting Hamiltonian from (\ref{Equ1}) into the form described by $\hat H$
in (\ref{Equ2a},\ref{Equ3}), are as follows:
One has 32 equations for nearest-neighbor contributions $m=1,2$, namely 16
for a fixed m
\begin{eqnarray}
- k^{p,p';\sigma,\sigma'}_{{\bf i},{\bf i}+{\bf x}_m}=\sum_{Q=A,B} (q^*_{Q,2m,p,\sigma} 
q_{Q,1,p',
\sigma'} + q^*_{Q,3,p,\sigma} q_{Q,6-2m,p',\sigma'}),
\label{Equx2}
\end{eqnarray}
and similarly one has 32 equations for the next nearest-neighbor contributions,
as previously 16 for a fixed $m= \pm 1$
\begin{eqnarray}
- k^{p,p';\sigma,\sigma'}_{{\bf i},{\bf i}+{\bf x}_2+m {\bf x}_1}=\sum_{Q=A,B} 
q^*_{Q,3+(1-m)/2,p,
\sigma} q_{Q,1+(1-m)/2,p',\sigma'}.
\label{Equx3}
\end{eqnarray}
Finally local (e.g. ${\bf r}=0$) contributions give rise to 16 equations
which can be written as
\begin{eqnarray}
&&-k^{p,p';\sigma,\sigma'}_{{\bf i},{\bf i}}[(1-\delta_{p,p'})+(1-\delta_{\sigma,\sigma'})
\delta_{p,p'} + \delta_{p,p'}
\delta_{\sigma,\sigma'}] +(\eta - U_p) \delta_{p,p'}\delta_{\sigma,\sigma'}
\nonumber\\ 
&&-[h^x-ik(\sigma)h^y] \delta_{p,p'}(1-\delta_{\sigma,\sigma'})
=\sum_{Q=A,B} \sum_{n=1,2,3,4} q^*_{Q,n,p,\sigma} q_{Q,n,p',\sigma'},
\label{Equx4}
\end{eqnarray}
where $k(\sigma)=\delta_{\uparrow,\sigma} -\delta_{\downarrow,\sigma}$.
One has here totally 80 non-linear equations, whose unknown are the 
32 numerical prefactors $q_{Q,n,p,\sigma}$
called ``block operator parameters'',
and the parameter $\eta$ entering in the ground state energy
($E_g=S_c$). The total number
of Hamiltonian parameters (taking into account all possible spin dependences 
as well) is 76, so a proper description for a real material can be provided.
But taking into account the conditions presented below (\ref{Equ1}) and used in
this description, besides SOI couplings and U, one remains with only 
10 $\hat H_{kin}$ parameters per one band in both ${\bf x}_1,{\bf x}_2$
directions.

\subsection{Solution of the matching equations}

In order to start the deduction of the exact ground states, first we should 
deduce the numerical prefactors $q_{Q,p,n,\alpha}$ of the block operators from 
(\ref{Equ2}) from the matching equations (\ref{Equx2}-\ref{Equx4}). Starting
this job, first we observe from (\ref{Equx2}-\ref{Equx4}) that all 
$q_{Q=A,p,n,\alpha}$ components can be given in function of the $q_{Q=B,p,n,\alpha}$ 
coefficients via the relation $q_{A,p,n,\alpha} = d_{n,\alpha} q_{B,p,n,\alpha}$, where
the coefficiens $d_{n,\alpha}$ have the expression
\begin{eqnarray}
d_{n,\alpha}=-(\frac{\delta_{\alpha,\uparrow}}{y} +\frac{\delta_{\alpha,\downarrow}}{x})
\delta_{n,1}-(\frac{\delta_{\alpha,\uparrow}}{v} +\frac{\delta_{\alpha,\downarrow}}{z})
\delta_{n,2} + (x^*\delta_{\alpha,\uparrow} +y^*\delta_{\alpha,\downarrow})\delta_{n,3} +
(z^*\delta_{\alpha,\uparrow} +v^*\delta_{\alpha,\downarrow})\delta_{n,4},
\label{Equx4a}
\end{eqnarray}
where $x,y,v,z$ are numerical prefactors. After this step it results that the
remaining $q_{B,p,n,\alpha}$ unknowns with $p=a$ can be given in term of the
$q_{B,p,n,\alpha}$ coefficients containing $p=b$ via the relation 
$q_{B,a,n,\alpha}=\alpha_n q_{B,b,n,\alpha}$, where one has for the numerical 
coefficients $\alpha_n$ the expression
\begin{eqnarray}
\alpha_n = \alpha_1 \delta_{n,1} + \alpha_2 \delta_{n,2} + \frac{\gamma_0}{
\alpha^*_1} \delta_{n,3} + \frac{\gamma_0}{\alpha^*_2} \delta_{n,4}
\label{Equx4b}
\end{eqnarray}
where $\gamma_0$ is an arbitrary real and positive parameter, 
while $\alpha_1,\alpha_2$ are
two further numerical prefactors. In this manner, up to (\ref{Equx4b}) only 8
unknown coefficients remain, namely $q_{B,b,n,\alpha}$ with $n=1,2,3,4$ and
$\alpha=\uparrow,\downarrow$. But it turns out that these eight unknown 
coefficients are interdependent, and all can be expressed in function of one
block operator parameter, namely $q_{B,b,n=1,\uparrow}$, via
\begin{eqnarray}
&&q_{B,b,1,\downarrow}=\frac{1}{w^*}q_{B,b,1,\uparrow}, \quad
q_{B,b,3,\downarrow}=-\frac{u^*}{w^*}q^*_{B,b,1,\uparrow}, \quad
q_{B,b,3,\uparrow}=\frac{|\alpha_1|^2}{\gamma_0} \frac{1}{u y x^*} q_{B,b,1,\uparrow},
\nonumber\\
&&q_{B,b,2,\downarrow}= \omega q_{B,b,2,\uparrow}, \quad
q_{B,b,4,\downarrow}=-\frac{\alpha^*_2}{\alpha^*_1}\frac{u^*y^*}{v^*w^*}q^*_{
B,b,2,\uparrow}, \quad
q_{B,b,4,\uparrow}=\frac{\alpha^*_2\alpha_1}{\gamma_0}\frac{\omega^*}{u y z^*}q^*_{
B,b,2,\uparrow}, 
\label{Equx4c}
\end{eqnarray}
where $|q_{B,b,2,\uparrow}|=|\mu||q_{B,b,1,\uparrow}|$, $\omega=[zwx^*(1+vy^*)]/
[vy^*(1+zx^*)]$, $|w|=(|u||y|\sqrt{\gamma_0})/|\alpha_1|, \sigma=yv^*$. 
Taking $\sigma, k, \phi_1, \phi_2$ as arbitrary parameters, one obtains three 
coupled equations in $X=vx^*,Z=vz^*, V=|v|^2$
\begin{eqnarray}
&&k\sigma^*(1+\sigma) = Z [k(1+\sigma) + X^* -\frac{|X|^2}{V}] +(V -X),
\nonumber\\
&&\frac{X-V}{1+X}e^{i\phi_1}=V\frac{\sigma^*-Z}{V+\sigma Z},
\nonumber\\
&&\frac{V+Z X^*}{X-Z}e^{i\phi_2}=V\frac{(1+\sigma^*)}{
\sigma-V}.
\label{Equ11}
\end{eqnarray}
from where, together with the $\sigma$ expression, the remaining unknown
$x,y,z,v$ parameters can be deduced, and based on them, starting
from the relation $k_2=f \: q_{B,b,n=1,\uparrow}$, where 
$k_2=|q_{B,b,n=2,\uparrow}|$ is a free
parameter, and $f \equiv f(x,y,z,v)$ is a known function, see
(\ref{Equ12}),  
\begin{eqnarray}
f=\frac{V(|\alpha_1|^2-\gamma_0)}{|y|^2(\gamma_0-|\alpha_2|^2)}
\frac{k(1+|y|^2)-(1+|x|^2)}{k(1+V)-(1+|z|^2)\frac{|v-x|^2}{|y-z|^2}},
\label{Equ12}
\end{eqnarray}
$q_{B,b,n=1,\uparrow}$ 
can be determined. Then, $q_{B,b,n=2,\uparrow}=|q_{B,b,n=2,\uparrow}|exp(i\theta_2)$, 
where $\theta_2$ is a free parameter, is given by $k_2\exp{i\theta_2}$.
The Hamiltonian parameters expressed in $k^{a,a}_{{\bf i},{\bf i}+
{\bf x}_1}$ units, enter in the 12 free parameters 
$k, Re(\sigma), Im(\sigma), \phi_1, \phi_2, Re(\alpha_1), Im(\alpha_1), 
Re(\alpha_2), Im(\alpha_2), \gamma_0, k_2, \theta_2$
of the solution presented above (e.g. $k^{a,b,\sigma,\sigma}_{{\bf i},{\bf i}+
{\bf x}_1}=\frac{2\alpha_2^*}{\alpha_1\alpha^*_2+\gamma_0}$,
$k^{b,a,\sigma,\sigma}_{{\bf i},{\bf i}+
{\bf x}_1}=\frac{2\alpha_1}{\alpha_1\alpha^*_2+\gamma_0}$, etc.).
We further note that when the presented solution appears, the relation
$k^{b,b,\sigma,\sigma}_{{\bf i},{\bf i}+{\bf r}}=(1/\gamma_0)k^{a,a,\sigma,\sigma}_{{\bf i},
{\bf i}+{\bf r}}$ fixes for all possible ${\bf r}$ [see the discussion following
(\ref{Equ1})] the magnitudes ratio of diagonal $\hat H_{kin}$ overlap elements 
from the two bands.

\section{The ground state wave functions }

The first deduced ground state wave function corresponding to the transformed
Hamiltonian from (\ref{Equ2a}) connected to the matching equations
(\ref{Equ3}-\ref{Equx4}) is of the form
\begin{eqnarray}
|\Psi_{1,g}\rangle = \prod_{\bf i} (\prod_{Q=A,B} \hat Q^{\dagger}_{\bf i})
\hat Q^{\dagger}_{1,{\bf i}} |0\rangle
\label{Equx5}
\end{eqnarray}
where $\prod_{\bf i}$ extends over all $N_{\Lambda}$ lattice sites, one has
$\hat Q^{\dagger}_{1,{\bf i}}= \sum_{\sigma} \alpha_{\sigma,{\bf i}} 
\hat c^{\dagger}_{b,{\bf i},\sigma}$, where $\alpha_{\sigma,{\bf i}}$ are numerical
prefactors, and $|0\rangle$ is the bare vacuum with no fermions present.
This $|\Psi_{1,g}\rangle$ solution corresponds to $3/4$ system filling.

The presented wave vector from (\ref{Equx5}) represents the ground state for
the following reason: a) As seen from (\ref{Equ2}) the block operators 
$\hat Q^{\dagger}_{\bf i}$ are linear combinations of canonical Fermi creation 
operators acting on the finite number of sites of the given block, consequently
the $\hat Q^{\dagger}_{\bf i}\hat Q^{\dagger}_{\bf i}=0$ equality is satisfied. Hence
the relation $\hat P_Q |\Psi_{1,g}\rangle=0$ automatically holds. Furthermore,
b) The $\hat P_{\bf i}$ positive semidefinite operators from the expression of 
the $\hat P_U$ operators in (\ref{Equ3}) (note that because of $U_b > 0$, also
$\hat P_U$ is a positive semidefinite operator) attain their minimum eigenvalue
zero when at least one b-electron is present on the site ${\bf i}$. Hence,
for the minimum eigenvalue zero of $\hat P_U$, at least one b-electron is needed
to be present on all lattice sites. But $\prod_{\bf i}\hat Q^{\dagger}_{1,{\bf i}}$ 
introduces a b-electron on each site, consequently also 
$\hat P_U |\Psi_{1,g}\rangle=0$ holds. As a summary of the above presented 
arguments, also for $\hat P=\hat P_Q+\hat P_U$ one has $\hat P |\Psi_{1,g}
\rangle=0$, i.e. $|\Psi_{1,g}\rangle$ represents the ground state. The uniqueness
of this ground state at $3/4$ system filling can also be demonstrated on the 
line of the Appendix 2 of Ref.[16].

We note that the ground state (\ref{Equx5}) can be extended also above $3/4$
system filling as follows:
\begin{eqnarray}
|\Psi_{2,g}\rangle = \prod_{\bf i} (\prod_{Q=A,B} \hat Q^{\dagger}_{\bf i})
\hat Q^{\dagger}_{1,{\bf i}} (\prod_{j=1}^{N_1} \hat c^{\dagger}_{b,{\bf k}_j,
\alpha_{{\bf k}_j}}) |0\rangle
\label{Equx6}
\end{eqnarray}
where $N_1 < N_{\Lambda}$, $\hat c^{\dagger}_{b,{\bf k},\alpha}$ is the Fourier 
transformed $\hat c^{\dagger}_{b,{\bf i},\alpha}$, $\alpha_{{\bf k}_j}$ being an 
arbitrary
spin projection for each ${\bf k}_j$, and  $\prod_{j=1}^{N_1}$ is taken over
$N_1$ arbitrary ${\bf k}_j$ values.  
The filling corresponding to (\ref{Equx6}) 
corresponds to $3/4 +N_1/N_{\Lambda}$ system filling. The demonstration of the 
ground state nature follows the line presented above in the case of
(\ref{Equx5}), and is based on the observation that the supplementary product
$(\prod_{j=1}^{N_1} \hat c^{\dagger}_{b,{\bf k}_j,\alpha_{{\bf k}_j}})$ not alters the
properties $\hat P_Q|\Psi_{z,g}\rangle=\hat P_U|\Psi_{z,g}\rangle=0$, for 
both $z=1,2$. 
 
In similar manner we have deduced ground state wave vectors also below system 
half filling. On this line one has
\begin{eqnarray}
|\Psi_{3,g}\rangle = \prod^{N_s}_j \hat C^{\dagger}_{j} |0\rangle
\label{Equx7}
\end{eqnarray} 
where $\hat C^{\dagger}_{j}$ represent block operators which on their turn are
linear combinations of $\hat c^{\dagger}_{p,{\bf i},\sigma}$ creation operators,
and must satisfy the 
anti-commutation relations $\{\hat Q_{\bf i},\hat C^{\dagger}_{j}\}=0$ for all
lattice sites ${\bf i}$, and both $Q=A,B$ values. The $j$ index here
denotes different (independent) $\hat C^{\dagger}_{j}$ terms. The number of 
carriers described by (\ref{Equx7}) is given by $N_s$. We underline that in the
case of the ground state (\ref{Equx7}) the starting Hamiltonian (\ref{Equ1})
is transformed in the expression
\begin{eqnarray}
\hat H = \hat P_{Q,1} + \eta N, \quad \hat P_{Q,1} = \sum_{\bf i}\sum_{Q=A,B} 
\hat Q^{\dagger}_{\bf i} \hat Q_{\bf i}.
\label{Equx8}
\end{eqnarray}
The corresponding matching equations (\ref{Equx2}-\ref{Equx4}) remain unaltered
in their right hand side, but their left hand side gains a minus sign, and 
supplementary, in (\ref{Equx4}) the renormation $\eta \to \eta + U_p$ emerges.
The energies $E_{n,g}$ corresponding to the ground states
$|\Psi_{n,g}\rangle$ for $n=1,2,3$ become
\begin{eqnarray}
E_{n,g}= [\eta (N + N_1\delta_{n,2}) -U_b N_{\Lambda} - \sum_{\bf i}\sum_{Q=A,B} 
d_{{\bf i},Q}](\delta_{n,1}+\delta_{n,2})+ \eta N_s \delta_{n,3}.
\label{Equx9}
\end{eqnarray}

\section{Summary and Conclusions}

We started with the observation that
surfaces and interfaces have a broad 
application spectrum in leading technologies, and such two dimensional systems 
have by their nature in the case of real materials potential gradients 
($\nabla V$) at their 
surfaces. These gradients are generating many-body spin-orbit coupling
(${\vec \sigma} \cdot (\nabla V \times {\vec k})$), which
even if small, produces essential effects since breaks the double spin 
projection degeneracy of each band. 
In applications, the correlated electron surfaces are often 
exposed to in-plane magnetic fields. Such processes, contrary to their 
importance, have not been analysed till now in exact terms 
mainly due to the non-integrable nature of the systems.
In this Letter we fill up
this gap by working out and describing a procedure which, using 
positive semidefinite operator properties, deduces exact results in 2D
for such systems. The technique merits attention since provide solution for the
matching equations consisting of 80 coupled non-linear complex-algebraic
equations. Extended results presenting how this technique works in
concrete cases will be published in a forthcoming detailed paper.    

{\bf Acknowledgements}:
N. K. acknowledges the support of ÚNKP-18-3 New National Excellence Program 
of the Hungarian Ministry of Human Capacities, while Z.G. of the
EU-funded Hungarian Grant No. EFOP-3.6.2-16-2017-00005,
and the Alexander von Humboldt Foundation.

\end{document}